\documentclass[manuscript]{emulateapj}
\usepackage{bm}

\def\nh {N_{\rm H}}
\def\xspec{{\sc xspec}}
\def\bb{{\sc bb}}
\def\ktbb{kT_{\rm bb}}
\def\ktw{kT_{\rm w}}
\def\kts{kT_{\rm s}}
\def\kte{kT_{\rm e}}

\def\bmc{{\sc bmc}}
\def\comptt{{\sc comptt}}
\def\compps{{\sc compps}}

\def\compst{{\sc compst}}
\def\comptb{{\sc comptb}}
\def\diskbb{{\sc diskbb}}

\def\gax {\ifmmode{_>\atop^{\sim}}\else{${_>\atop^{\sim}}$}\fi}  
\def\pl{{\sc pl}}
\def\gaussian{{\sc gaussian}}
\def\chiq{$\chi^2$}
\def\cm2{cm$^{-2}$}
\def\s1{s$^{-1}$}

\def\wabs{{\sc wabs}}
\def\sax{\mbox{{\it BeppoSAX}}}
\def\integral{\mbox{{\it INTEGRAL}}}
\def\xte{{\it RXTE}}

\shorttitle{A NEW COMPTONIZATION MODEL FOR NEUTRON STAR LMXBS}
\shortauthors{FARINELLI ET AL.}

\begin{document}

\title{A new Comptonization model for low-magnetized accreting neutron stars in low mass X-ray binaries}

\author{Ruben Farinelli\altaffilmark{1}, Lev Titarchuk\altaffilmark{1,2,3}, Ada Paizis\altaffilmark{4} \&  
Filippo Frontera\altaffilmark{1,5}}
\altaffiltext{1}{Dipartimento di Fisica, Universit\'a di Ferrara, via Saragat 1, 44100, Ferrara, Italy; 
farinelli@fe.infn.it}
\altaffiltext{2}{George Mason University/CEOSR/US Naval Research Laboratory, Code 7655, Washington DC 20375; Lev.Titarchuk@nrl.navy.mil}
\altaffiltext{3}{National Aeronautics and Space Administration, Goddard Space Flight Center (NASA/GSFC), Greenbelt, MD 20771, USA}
\altaffiltext{4}{INAF-IASF, Sezione di Milano, Via Bassini 15, 20100, Milano, Italy}
\altaffiltext{5}{INAF-IASF, Sezione di Bologna, Via Gobetti 1, 40100, Bologna, Italy}

\begin{abstract}

We developed a new model for the X-ray spectral fitting \xspec\ package which takes into account the effects of both thermal
and dynamical (i.e. bulk) Comptonization.
The model consists  of two components: one  is the direct blackbody-like emission due to seed photons which are not subjected to effective Compton scattering, while the other one is a convolution of the Green's function of the energy operator with a blackbody-like 
seed photon spectrum. 
When combined thermal and bulk effects are considered, the analytic form of the Green's function may be obtained as a solution of
the diffusion Comptonization equation.
Using data from the \sax, \integral\ and \xte\ satellites, we test our model on the  spectra 
of a sample of six persistently low magnetic field bright neutron star Low Mass \mbox{X-ray} Binaries, covering three different  spectral states.
Particular attention is given to the transient powerlaw-like hard  \mbox{X-ray} ($\ga 30$\,keV) tails
that we interpret in the framework of the bulk motion Comptonization process.
We show that the values of the  best-fit $\delta$-parameter, which represents the importance of bulk with 
respect to thermal  Comptonization, can be physically meaningful and can at least 
qualitatively describe the physical conditions of the environment in the innermost 
part of the system. Moreover, we show that in fitting the thermal Comptonization spectra to
the \mbox{X-ray} spectra of these systems, the best-fit parameters of our model are 
in excellent agreement with those of \comptt, a  broadly used and well established \xspec\ model. 
\end{abstract}


\maketitle

\section{Introduction}
\label{introduction}

The study of the transient hard \mbox{X-ray} tails in neutron star (NS) Low Mass \mbox{X-ray} Binaries (LMXBs)
has recently received a strong theoretical impulse. The first approach has been mainly 
phenomenological, with the hard tails simply fitted by a powerlaw (\pl) model (see review 
by Di Salvo \& Stella 2002).
Subsequently, using 0.4-120\,keV \sax\ data, Farinelli et al. (2005, hereafter F05) tried a more 
physical approach, describing 
the hard tail of \mbox{GX~17$+$2} by means of Comptonization of soft 
($\sim$ 0.6\,keV) seed photons off a hybrid (thermal plus non-thermal) electron population.
More recently, Paizis et al. (2006, hereafter P06), using  long-term time-averaged 
\integral/IBIS spectra ($>$ 20\,keV), confirmed the presence of hard \mbox{X-ray} emission in the Z sources \mbox{Sco X--1}, 
\mbox{GX~5--1}, \mbox{GX~17$+$2}, \mbox{GX~340$+$0}, \mbox{Cyg~X--2}, and they also discovered it  in the atoll source
\mbox{GX~13$+$1}, which, in fact, is characterized by a stable Z-like \mbox{X-ray} continuum (thermal component 
with low electron temperature $\kte$ and high optical depth $\tau$). 

Motivated by observations of the intermediate and soft states of black 
hole candidates (BHCs), P06 proposed that  bulk motion Comptonization can be responsible  for the formation  of the 
well-known hard \mbox{X-ray} tails in NSs observed in these sources as well.
In their analysis, they used the Bulk Motion Comptonization (\bmc) model in \xspec, whose 
description is reported in Titarchuk et al. (1996, 1997, hereafter TMK96 and TMK97, respectively). 
Moreover, merging their results with those of Falanga et al. (2006, hereafter F06) on the atoll source 
4U~1728--34 (GX~354--0), P06
identified four main spectral states for NS LMXBs (see Fig. 4 in P06): \emph{hard/\pl\ state}  (\mbox{H~1750--440} and \mbox{H~1608--55}), \emph{low/hard state} (\mbox{GX 354--0}), \emph{intermediate state} (e.g., \mbox{GX 5--1}) and \emph{very soft state} (e.g., GX~3$+$1). In their picture, the 
relative contribution of thermal and bulk Comptonization (TC and BC, respectively), in turn related 
to the accretion rate $\dot{M}$ and the radiation pressure from the NS, is the main 
parameter explaining both the different spectral states among several sources and the 
spectral evolution of a single source.
However, the lack of data below 20\,keV prevented P06 from drawing more stringent 
conclusions on the accretion geometry of the systems and on the physical parameters.

Re-analyzing broad-band \sax\ data, Farinelli et al. (2007, hereafter F07) applied the 
\bmc\ model to  \mbox{GX~17$+$2}. They found that the source spectrum can be described by a blackbody (\bb)-like emission,
plus an unsaturated  TC component plus a \pl. 
The TC component, fitted with the \comptt\ model (Titarchuk 1994, hereafter T94), is suggested to come from the
region (called transition layer, TL) between the Keplerian accretion disk and the NS surface, 
while the \pl-like emission is attributed to BC of the NS seed photons on the 
in-falling material, which is in turn the innermost part of the TL itself.
This accretion scenario proposed by F07 is supported by results obtained by numerical solutions of 
 the equation for the radial velocity  profile $v_{\rm R}$ in accretion disks
(Titarchuk \& Farinelli 2008, hereafter TF08, in preparation), where the quasi free-fall behaviour 
of $v_{\rm R}$ inside the TL from some  radius $R_{\rm ff}$ to the NS surface, is unambiguously shown.

It is thus evident that BC is gaining strong theoretical and 
observational support in the study of LMXB spectral evolution. 
This is true also for the case of accretion  powered \mbox{X-ray} pulsars for which a new theoretical model based on  
TC and BC, occurring in the accreting shocked gas, 
was recently presented (Becker \& Wolff, 2007, hereafter BW07).
In the BW07 Comptonization model the effects of the strong magnetic field ($B \sim 10^{12}$ G) are very important, as the
accreting gas column is channelized at the NS magnetic poles and the presence of this field also has important consequences for the photons propagating through the plasma. 

From the observational point of view, the energy index $\alpha$ and the cutoff energy $E_{\rm c}$ are only parameters which can be determined by fitting  the observable spectral shapes to the Comptonization models; additional assumptions have to be adopted however in order to relate them to the physical parameters of the particular models. 

However in this Paper we concentrate  in the study of sources where the magnetic field is low, so that it has second-order effects with respect to radiation pressure in driving the accretion process.
For a such class of low-magnetized compact objects, in the current \xspec\ package it is available the \bmc\ model, which was used, as mentioned above, by P06. It is in fact a \emph{general Comptonization} 
model, i.e. widely applicable, given that the output emerging Comptonized spectrum is given by the 
convolution of the analytical approximation of the \emph{Green's  function} (response of the system to  injection of a  monochromatic line at energy $ E_0$)
with a \bb\  seed photon spectrum, regardless of the  Comptonization process, i.e. only due to thermal or combined thermal and  bulk effect.

The \bmc\ model parameters are the \bb\ seed photons color temperature ($\ktbb$) and  normalization, 
the energy index $\alpha$ of the Comptonized spectrum and the \emph{illumination factor} 
$\log(A)$. 
The information on the efficiency of the Comptonization is carried-out by the 
index $\alpha$: the lower $\alpha$, the higher the energy exchange 
from hotter electrons to softer seed photons. As we mention above, $\alpha$  
does not specify which kind of Comptonization process produces it, but it is simply 
related to an observable quantity in the photon spectrum of the data.
The Green's function of the model is approximated by a broken \pl, a condition 
that holds when the average energy of the seed photons is much lower than the average energy $E_{\rm av}$ of the
plasma (Sunyaev \& Titarchuk 1980, hereafter ST80, and TMK97).
The \bmc\ model however lacks the electron recoil-effect term in the Green's function, which is 
responsible for the rollover in the spectrum (the energy of the rollover, depending on the plasma
temperature and on the first and second order of the bulk velocity,  at  about $E_{\rm av}$).

Thus the Comptonized part of the emerging spectrum is in the \bmc\ model always a \pl\ with no  cutoff. 
However the lack of the energy cutoff  in the model  becomes critical in fitting of  the TC spectra of LMXBs in their high-luminosity state, when the rollover is well observed around 10\,keV.

Motivated by these limitations in the currently available version of the \bmc\  model for \xspec\ 
and by the recent observational results on LMXBs (see above), we developed a new model for \xspec\ which 
can be considered as an extension and completion of \bmc.
In Section \ref{description} we summarize the main results of the Bulk Motion Comptonization theory and 
 show the mathematical background of our model. 
In Section \ref{sources} we show the results of the \mbox{X-ray} spectral fitting using our model on a sample of six 
persistently bright NS LMXBs. In Section \ref{discussion} we discuss the main results and the limits 
of applicability of the model 
itself, with a breif theoretical comparison to the model of BW07,  and in Section \ref{conclusion} we draw our conclusions.

\section{Description of the model}
\label{description}
\subsection{Solution of the Comptonization equation for bulk flow of non-relativistic electrons}
\label{komp_equation}
The Comptonization  equation  (a particular type of  Fokker-Planck equation, see e.g. T94) describes an evolution of the photon field 
due to  Compton scattering of photons off  non-relativistic thermal electrons.
It is obtained as the approximation of the full kinetic Boltzmann equation when the photon field is
almost isotropic, the average number of scatterings is high and the average energy exchange per scattering is small, namely $\Delta E/E \ll 1$.
When the divergence of the velocity field $\nabla \cdot \vec{v_{\rm b}}$  of the electrons is zero (namely when electrons
are subjected to pure Brownian motion) only TC contributes to the total emerging
spectrum (ST80). On the other hand, if the plasma flow is subjected to inward bulk motion ($\nabla \cdot \vec{v_{\rm b}} < 0$), the latter provides a further channel to Comptonization.
The Comptonization equation including the bulk motion term was first derived and introduced by  Blandford \& Payne (1981). 
A slightly modified version of the equation, which includes  the second order bulk  term ${V{\rm b}}^2$   is presented in TMK97. 
The equation for the photon occupation number $n(x,\tau)$ in the case  of  a free-fall velocity profile ($v_r \propto r^{-1/2}$) reads as

\begin{flushleft}
\begin{eqnarray}
\tau { {\partial^2 n} \over {\partial \tau^2} }
- \left( \tau + {3 \over 2} \right) 
{ {\partial n} \over {\partial \tau} } = 
{1 \over 2} x { {\partial n} \over {\partial x} }
- { 1 \over {2\delta_b} } { 1 \over {x^2} } 
{ \partial \over {\partial x} } \nonumber \\ 
\times \left[ x^4 \left( f_b^{-1}n + { {\partial n} \over {\partial x} } 
\right) \right]
- {{\dot m} \over {2}}  
{ j \over {\kappa c} }, ~
\label{kompaneets}
\end{eqnarray}
\end{flushleft}
where $x\equiv E/\kte$ is a dimensionless energy,   
$\Theta \equiv \kte/m_{\rm e}c^2$ is a dimensionless electron temperature, ${\dot m}$ is the accretion rate in units of the  Eddington 
accretion rate,
 $\kappa$ is the inverse  of scattering free path of electron   and $\tau$ is the  optical depth.
The term $j$ in  equation (\ref{kompaneets}) is a function of $x$ and $\tau$ and represents the seed photons source function and $\delta_b^{-1} \equiv \delta^{-1}f_b(\tau)$, $f_b(\tau)=1+(v_{\rm b}/c)^2/(3\Theta)$.

The quantity of interest  in the above equation is the bulk parameter $\delta$, which is derived in TMK97  as
\begin{eqnarray}
\delta \equiv \displaystyle{{<E_{\rm bulk}>} \over {<E_{\rm th}>}}=\displaystyle{{\sqrt{(1-\ell)}\over{\Theta \dot{m}}}},
\label{delta}
\end{eqnarray}
where $\ell \equiv L/L_{Edd}$ is the fractional Eddington luminosity impinging on the flow.

It is worth noting that in equation~(\ref{kompaneets}), the term  $f_b^{-1}n$ depends  on $\tau$ through the bulk term $v^2_b$. 
Given that $f_b^{-1}n$ depends  on $\tau$  it is possible to find an analytical solution of the equation (using
the separation of variables method) only neglecting the $v_{\rm b}^2/c^2$ term, namely by setting $f_b$=1.
In this case the Green's function  $G(x,x_0)$=$N_{\rm G}(x_0) \hat G(x,x_0)$ (see also Eqs. [B4]-[B8] in TMK97) with $\hat G(x,x_0)$ given by

\begin{displaymath}
\hat{G}(x, x_0)=\displaystyle{{e^{-x}}\over{x_0~\Gamma (2\alpha+4+\delta)}}  
\end{displaymath}

\begin{equation}
\times \cases{_1F_1(\alpha,4+2\alpha+\delta,x)~\left({x \over {x_0} }\right)^{\alpha + \delta +3}J (x_0,\alpha,\omega), ~x \le x_0;\cr
 _1F_1(\alpha,4+2\alpha+\delta,x_0)\left({x \over {x_0} }\right)^{-\alpha}~~~ J(x,\alpha,\omega),~~~x \ge x_0.}
\label{green}
\end{equation}

Here $\alpha$ is energy index, $\omega$=$\alpha + \delta +3$ and $x$,   $x_0$ represent the dimensionless  scattered  and  injected photon energies, respectively, namely $x \equiv E/\kte$,  $x_0 \equiv E_0/\kte$. 
This is in fact the \emph{complete} form of the Green's function for the Comptonization equation, while in equation (22) of TMK97 it is reported the asymptotic case of low-energy ($x_0 \ll$ 1) monochromatic line injection.
The dependence of  spectral energy index $\alpha$ on the model physical parameters  $\Theta$, $\tau$ and  $\delta$ is presented in equations (7) and (23) of TMK96 and TMK97, respectively. 
We note that the dependence of $\alpha$ on  $\Theta$ and  $\tau$ in  equation (\ref{green}) is implicit through the presence of the first eigenvalue of the space diffusion operator.

In  equation~(\ref{green}), $_1F_1$  is a confluent hyper-geometric 
function (Abramowitz \& Stegun 1970), and $J(x,\alpha, \omega)$ is an  integral function 
expressed through the formula
\begin{eqnarray}
J(x,\alpha,\omega)=\int_0^\infty e^{-t} (x+t)^{\omega}~t^{\alpha-1}~dt.
\label{int_function}
\end{eqnarray}

Given that Comptonization conserves the photon number, this photon conservation law can be rewritten as the integral equation  for the Green's function:
\begin{equation}
\int_0^\infty {{G(x,x_0)} \over {x}} dx = {{1} \over {x_0}}.
\label{green_norm}
\end{equation}

Thus the normalization factor $N_{\rm G}(x_0)$ in equation (\ref{green}), which is dependent of the injected line energy $x_0$ reads as

\begin{equation}
N_{\rm G}(x_0)= \left[x_0 \int_0^\infty {{\hat G(x,x_0)} \over {x}}dx\right]^{-1}.
\label{num_norm}
\end{equation}
When the energy of the injected line is much smaller than the plasma energy, i.e. when  $x_0, ~x  \ll \omega $, the Green's  function is a broken \pl\ and  $N_{\rm G}(x_0) \approx \alpha (\alpha + \delta +3)$ (see TMK97). 

The emergent Comptonized  spectrum is given by the convolution  of the Green's  function with  $S(x_0)$ if the input spectrum, instead of being a simple monochromatic line, has a continuum shape $S(x_0)$. Let us consider seed photons distributed according to the law
\begin{equation}
S(x_0)=\displaystyle{ {C_{\rm N} x_0^\gamma} \over {e^{T_{\rm e}x_{\rm 0}/T_{\rm s}}-1}},
\label{source_func}
\end{equation}
where $T_{\rm e}$  and $T_{\rm s}$ are the electron and seed photon temperatures, respectively, and $C_{\rm N}$ is the normalization factor. This is evidently a blackbody  spectrum if  $\gamma$=3. 
The  Comptonization spectrum is thus given by

\begin{flushleft}
\begin{displaymath}
f(x) =S(x_0) \ast G(x,x_0)=\displaystyle{J(x,\alpha,\omega)~e^{-x}~x^{-\alpha} \over { \Gamma (2\alpha+4+\delta)}} \times
\end{displaymath}
\begin{displaymath}
\displaystyle{{\int_0^x N_G(x_0) _1F_1(\alpha,4+2\alpha+\delta,x_0)~x_0^{\alpha-1}~{{x_0{^\gamma}} \over {e^{T_{\rm e} x_{\rm 0}/T_{\rm s}}-1}}}dx_0}
\end{displaymath}
\begin{eqnarray}
\displaystyle{+~{e^{-x}~x^{\alpha+\delta+3} \over {\Gamma (2\alpha+4+\delta)}}~~_1F_1(\alpha,4+2\alpha+\delta,x)}\times \nonumber \\
\displaystyle{ {\int_x^\infty N_G(x_0)~J(x_0,\alpha, \omega){1 \over {x_0}^{\alpha+\delta+4}}~ {{x_0^\gamma} \over {e^{T_{\rm e} x_{\rm s}/T_{\rm 0}}-1}}} dx_0}.
\label{convolution}
\end{eqnarray}
\end{flushleft}

\begin{figure}
\includegraphics[width=0.7\linewidth,angle=-90]{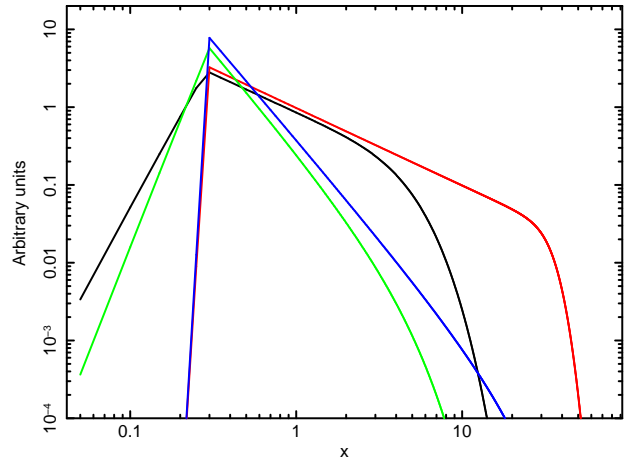}
\caption[]{The Green's functions $G(x,x_0)$ for monochromatic line with input dimensionless energy $x_0$=0.3 and 
different values of $\alpha$ and $\delta$: $\alpha$=1, $\delta$=0 ({\it black}); 
$\alpha$=1, $\delta$=30 ({\it red}); $\alpha$=2.5, $\delta$=0 ({\it green}); $\alpha$=2.5, 
$\delta$=30 ({\it blue}). The increase of the cutoff energy due to bulk effect 
is clearly visible.}
\label{green_plot}
\end{figure}

Given that the two integrals in equation (\ref{convolution}) do not have analytic presentation, we have  to  calculate them  numerically. This step is performed using the \emph{Gnu Scientific Library}\footnote{The 
Gnu Scientific Library is freely available at the web site www.gsl.org} package.
Note also that in the same equation the functions  $N_G(x_0)$ and $J(x_0, \alpha, \omega)$ are themselves integrals (see Eqs. [\ref{int_function}] and [\ref{num_norm}] ) and thus should be preliminary calculated before putting them into the integral.
This embedded integral operation is however very high CPU-time consuming, and it makes the whole convolution process very slow, for \xspec\ running. 

To avoid this problem, the integral function in equation (\ref{int_function}) is analytically  solved at each
integration step using the \emph{steepest descend method}, described in  Appendix C of TMK97.
This method provides very high accuracy (within few percent) for high  $\delta$-values ($\ga$ 10), 
while when $\delta$ is close to zero the discrepancy may reach 10\% (depending on the corresponding $\alpha$ and $x$-values).
To better approximate the integral value when  $\delta$ is low,  we produced a three-dimensional correction matrix of $\delta$, $\alpha$ and $x$-values (with $\delta$ in the interval 0-10)  which  provides a correction factor between the numerical and analytical results of integral (\ref{int_function}). The correction factor is computed using linear interpolation   among the  $\delta$, $\alpha$ and $x$-bounds of the grid.  The same approach is adopted to compute the normalization constant $N_G(x_0)$; for $\delta \ga 10$ we simply use the value $\alpha(\alpha+\delta+3)$, while for lower $\delta$-values we
use an interpolation grid. A subsequent numerical check reveals that the photon number is conserved with high accuracy (within few percent) over a very large spanning of the $\alpha$ and $\delta$-values.

\subsection{The free parameters of \comptb}
\label{comptb}
Following the same approach of the \bmc\ model (TMK96, TMK97), we structure our model as being the sum of two components: 
one represents the emission of the soft seed photons  which are not affected  by a  noticeable up-scattering in the 
plasma cloud, while the other one is a result of efficient Comptonization.
The emergent spectrum, similarly to \bmc,  can thus be described as

\begin{equation}
F(E)= \frac{C_{\rm N}}{A+1} (BB + A \times BB \ast G),
\label{flux}
\end{equation}
where $BB \equiv S(x_0)$ (see Eq. [\ref{source_func}])  and  the convolution   $BB \ast G$ is given by equation (\ref{convolution}). The free  parameters of our model are the seed photon  temperature $\kts$, the index 
$\gamma$ (if $\gamma$=3 seed photons have pure \bb\ spectrum), the Green's function energy index $\alpha$ (see Eq. [\ref{green}]), the bulk parameter $\delta$, the temperature of the electrons $\kte$, the 
 illuminating factor log(A) and the normalization constant $C_{\rm N}$. 
The latter is chosen is such a way that
when $A \rightarrow 0$ (and $\gamma$=3) the model simply reduces to the standard \bb\ model for \xspec.
We note also that $C_{\rm N}$ must be multiplied by the factor $0.5 \times 10^{23} (kT_{\rm e}/\textrm{keV})^4$ in order
to express the flux in physical units (erg\,cm$^{-2}$\,s$^{-1}$\,keV$^{-1}$).
In equation~(\ref{flux}) the factor  $1/(1+A)$ is the fraction of the seed photon radiation 
directly seen by the observer, whereas the factor $A/(1+A)$ is the fraction of 
the seed photon radiation up-scattered by the Compton cloud. 

As it can be clearly  seen, we expand the parameter set with respect to \bmc.\  In principle, this allows  one  to extract a more detailed physical information from the observed spectra. In Figure \ref{green_plot} we demonstrate the bulk motion effect presenting the Green's functions  
$G(x,x_0)$ for a couple of $\alpha$-values (1 and 2.5) and two different $\delta$-values (0 and 30) for each $\alpha$.
The cutoff energy is higher for greater $\delta$-values (see red and blue with 
respect to  black and green curves  related to $G(x,x_0)$ for which $\delta$=0).  
Note that the photon number $N_G(x_0)$  is the same for all $G(x,x_0)$. 
We want also to emphasize however,  that our model  \emph{is not specific} to bulk motion: it is evident indeed from equations (\ref{green}) and  (\ref{flux}) that setting $A \gg 1$ and $\delta$=0, a pure TC spectrum is reproduced, like the mostly used models \compst\ (ST80) and \comptt\ (T94).

It is also   important to point out  the difference between 
the standard ones and our Comptonization model.  In our case the code provides, as the best-fit parameters, the electron temperature  $\kte$ and the energy spectral index $\alpha$  instead  of $\kte$ and optical depth $\tau$ as  in   \compst\  and \comptt.
Thus, once $\kte$ and $\alpha$ are provided, it is possible to infer the Thomson optical depth $\tau$ using the relations reported in  Titarchuk \& Lyubasrkij (1995, hereafter TL95) for both spherical and slab geometry of the Comptonizing plasma.
For example, in the non-relativistic limit the following equation holds:
\begin{eqnarray}
\alpha=-\displaystyle{3 \over 2} + \sqrt{\displaystyle{9 \over 4} + \displaystyle{{\pi^2~m_{\rm e} c^2} \over {{\cal C}_{\tau}~kT_{\rm e}~(\tau+2/3)^2}}}
\label{alpha_kte_tau}
\end{eqnarray}
where $\tau$ is the optical radius of the sphere and optical half-thickness of the slab for spherical and plane geometries, respectively, 
 ${\cal C}_{\tau}$=3 for a sphere and ${\cal C}_{\tau}$=12 for a slab.
When bulk is present ($\delta >$ 0), the relation between $\alpha$, $\delta$ and $\kte$ is not straightforward and must be calculated numerically (see TMK97).

\begin{figure*}[!th]
\begin{center}
\includegraphics[width=16cm, height=14cm]{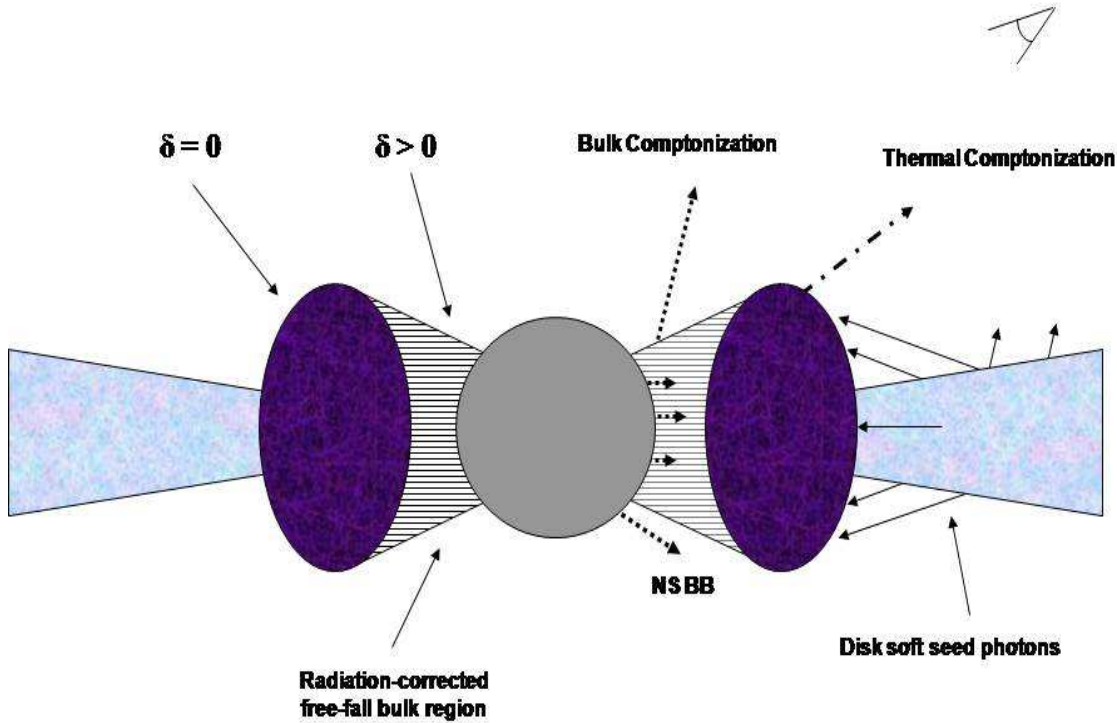}
\caption{Schematic view of the proposed geometry for thermal and bulk Comptonization regions in
LMXBs hosting a NS with \pl-like emission at high energies.  The thermal Comptonization spectrum (thermal \comptb) originates 
in the outer part of the TL region, whereas the thermal plus bulk Comptonization
spectrum (thermal plus bulk \comptb) arises in the innermost part of the TL where the  
\bb-like (NS) seed photons are (thermally and dynamically) Comptonized by the
in-falling material. The radial extension of the two Comptonization regions is, in fact, rescaled for clarity, whereas numerical calculations show that they are of order $R_{\rm NS}$.}  
\label{geometry}
\end{center}
\end{figure*}

\begin{figure*}[!th]
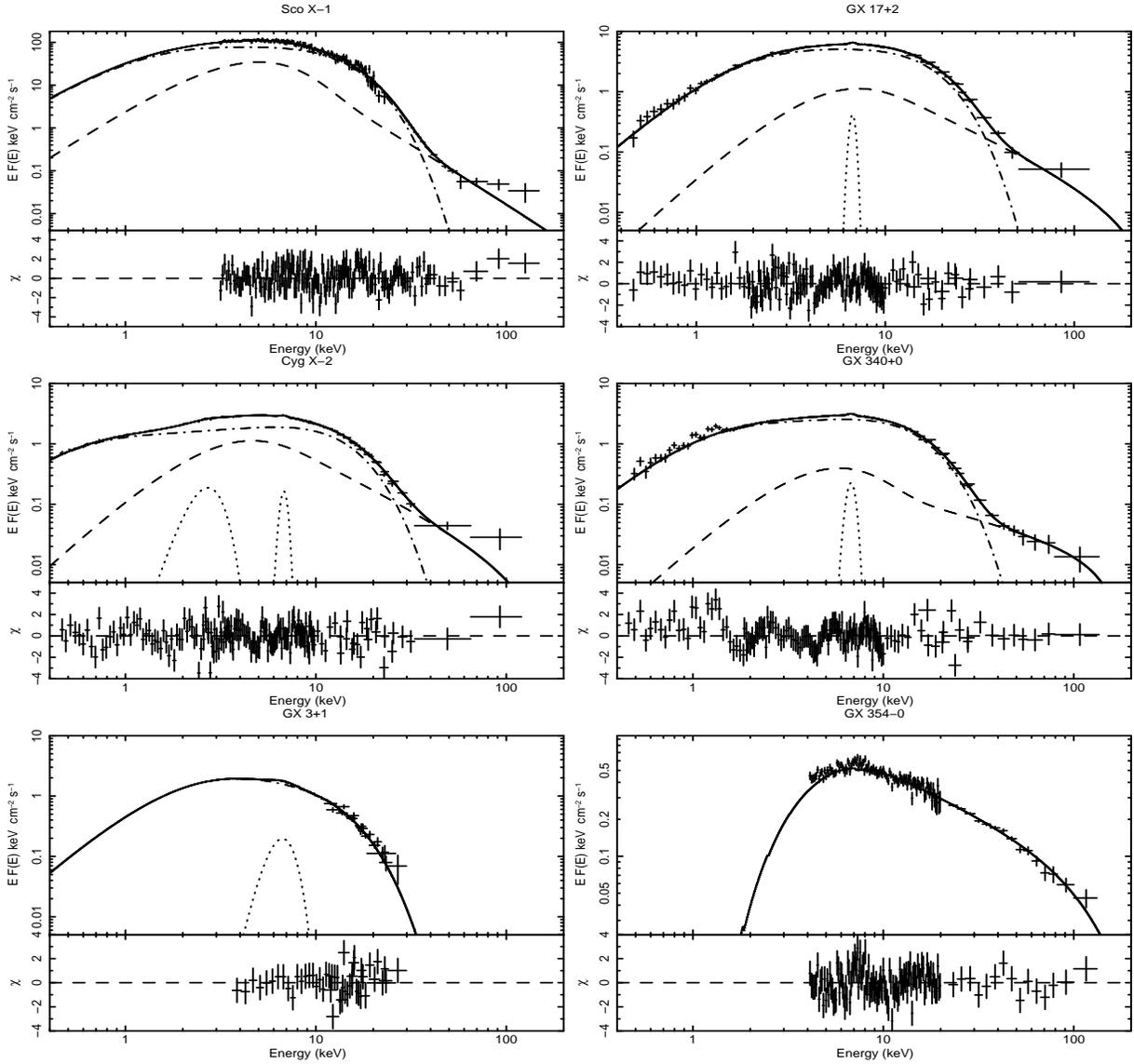

\begin{center}
\includegraphics[width=5cm, height=8cm, angle=-90]{f3a.eps}
\includegraphics[width=5cm, height=8cm, angle=-90]{f3b.eps}
\includegraphics[width=5cm, height=8cm, angle=-90]{f3c.eps}
\includegraphics[width=5cm, height=8cm, angle=-90]{f3d.eps}
\includegraphics[width=5cm, height=8cm, angle=-90]{f3e.eps}
\includegraphics[width=5cm, height=8cm, angle=-90]{f3f.eps}
\caption{Unabsorbed EF(E) spectra of the six studied NS LMXBs, superposed with the best-fit models 
({\it solid line}, see Table  \ref{fit_comptb}) and residuals between the data and 
the models in units of $\sigma$. For \mbox{Sco X--1}, \mbox{GX~17+2}, \mbox{Cyg~X--2} and \mbox{GX~340$+$0}, 
the model consists of two \comptb\ components: one represents pure thermal Comptonization 
({\it dotted-dashed line}), the other one represents  thermal plus bulk Comptonization ({\it long-dashed line}). For GX 3+1 and \mbox{GX 354--0} the model is a simple \comptb\ with pure thermal Comptonization ({\it solid line}). 
Gaussian emission lines, when required by the data,  are also shown ({\it dotted line}).} 
\label{efe}
\end{center}
\end{figure*}

\begin{table*}
\begin{center}
 \caption[]{Best-fit parameters to the analyzed sources using \comptt\ as thermal Comptonization model. For \mbox{GX~17$+$2}, Cyg~X--2 and GX~340$+$0 additional \bb\ and \pl\ components are required, while for Sco X--1 only a \pl\ is required to describe the high-energy part of the spectrum. 
For  GX 3+1 and GX 354--0 the continuum model is a simple \comptt. Emission lines, where observed, are modeled with a simple \gaussian\ model.  Parameters within square brackets are kept  frozen in the fit. Photoelectric absorption is computed using the \wabs\ model in \xspec. Errors are computed at 90\% confidence level for a single parameter.}
\begin{tabular}{ccccccc}
\noalign{\smallskip}
\hline
\hline
\noalign{\smallskip}
Parameter          &   Sco~X--1	   &	\mbox{GX~17$+$2}         &    Cyg~X--2   &   GX~340$+$0   & GX~3+1 &  GX 354--0\\
\noalign{\smallskip}
\hline
\noalign{\smallskip}
$N_{\rm H}^{(a)}$ &  [0.18] & 2.43$^{+  0.22}_{-  0.26}$  & 0.29$^{+  0.05}_{- 0.04}$  & 7.92$^{+  0.12}_{-  0.20}$   & [1.7]  & [3.0]  \\
\noalign{\smallskip}

\hline
\noalign{\smallskip}
\multicolumn{7}{c}{{\sc bb}}   \\
\noalign{\smallskip}
\hline
\noalign{\smallskip}
$kT_{\rm bb}^{(b)}$ (keV) &  --   &   1.57$^{+  0.09}_{-  0.08}$  &  1.18$^{+  0.03}_{-  0.04}$   &  1.40$^{+  0.04}_{-  0.05}$ & -- & --  \\

\noalign{\smallskip}
$R_{\rm bb}^{(c)}$ (km) &  --   &   4.6$^{+  0.5}_{-  0.5}$   &  8.0$^{+  0.5}_{-  0.5}$   &  6.5$^{+   0.5}_{-   0.5}$  & --  &  --  \\
\noalign{\smallskip}
\hline
\noalign{\smallskip}
\multicolumn{7}{c}{{\sc comptt}}   \\
\noalign{\smallskip}
\hline
\noalign{\smallskip}
$kT_{\rm w}^{(d)}$ (keV) & 1.08$^{+  0.07}_{-  0.07}$  & 0.55$^{+  0.02}_{-  0.02}$   & 0.17$^{+  0.03}_{-  0.03}$  & 0.43$^{+  0.09}_{-  0.33}$ & 0.62$^{+  0.08}_{-  0.11}$  &  1.18$^{+  0.04}_{-  0.04}$   \\
\noalign{\smallskip}

$kT_{\rm e}$ (keV) &  2.75$^{+  0.03}_{-  0.03}$    &    3.37$^{+  0.07}_{-  0.06}$   &  2.79$^{+  0.04}_{-  0.06}$  & 2.80$^{+  0.04}_{-  0.04}$ & 2.54$^{+  0.10}_{-  0.08}$ &  31.1$^{+ 13.8}_{-  7.1}$  \\
\noalign{\smallskip}
$\tau$ &  4.9$^{+  0.2}_{-  0.2}$ & 5.9$^{+   0.2}_{-   0.2}$  &  6.9$^{+  0.5}_{-  0.2}$  & 9.2$^{+  4.8}_{-  1.0}$  &5.3$^{+  0.27}_{-  0.30}$  &  1.5$^{+  0.6}_{-  0.6}$ \\
\noalign{\smallskip}
\hline
\noalign{\smallskip}

\multicolumn{7}{c}{{\sc powerlaw}} \\
\noalign{\smallskip}
\hline
\noalign{\smallskip}

$\Gamma$  &   4.15$^{+  0.08}_{-  0.10}$   &    2.75$^{+  0.15}_{-  0.23}$   & 2.52$^{+  0.21}_{-  0.21}$  & 3.41$^{+  0.05}_{-  0.05}$ & --  & -- \\
\noalign{\smallskip}
$N_{\rm pl}^{(e)}$ & 524$^{+152}_{-151}$  & 1.75$^{+  1.19}_{-  1.04}$  &  0.33$^{+  0.30}_{-  0.17}$    & 8.4$^{+  1.4}_{-  1.8}$  &  -- & -- \\
\noalign{\smallskip}
\hline

\noalign{\smallskip}

\multicolumn{7}{c}{{\sc gaussian}} \\
\noalign{\smallskip}
\hline
\noalign{\smallskip}

$E_{\rm l}$ (keV) & -- &   6.70$^{+  0.05}_{-  0.05}$    &  6.78$^{+  0.10}_{-  0.09}$  & 6.79$^{+  0.07}_{-  0.07}$  & [6.4] & -- \\
\noalign{\smallskip}
$\sigma_{\rm l}$ (keV) &   --    & 0.24$^{+  0.11}_{-  0.09}$  &  0.21$^{+  0.15}_{-  0.15}$   & 0.29$^{+  0.14}_{-  0.14}$ & 1.02$^{+  0.31}_{-  0.30}$  & -- \\
\noalign{\smallskip}
$I_{\rm l}^{(f)}$ &  --  & 5.1$^{+ 1.0}_{-   1.1}$   &  2.0$^{+0.1}_{-0.1}$    &  3.3$^{+  0.8}_{-  0.8}$ & 13$^{+5}_{-5}$   & -- \\
\noalign{\smallskip}
$EW_{\rm l}$ (eV)  &   -- & 43$^{+1}_{-14} $  & 27$^{+13}_{-7}$     &   48$^{+16}_{-10}$  & 357$^{+104}_{-126}$ & -- \\
\noalign{\smallskip}
\hline
\noalign{\smallskip}

${\chi^{2}}$/dof & 212/167 &  155/144   &    196/152     & 174/137   & 50/42   &  151/131 \\
\noalign{\smallskip}
\hline
\noalign{\smallskip}
\multicolumn{7}{l}{$^{\rm a}$ Equivalent hydrogen column in units of 10$^{22}$ cm$^{-2}$.} \\
\multicolumn{7}{l}{$^{\rm b}$ Temperature of the  blackbody spectrum.} \\
\multicolumn{7}{l}{$^{\rm c}$ Computed assuming isotropic emission and given distance for each source (see text).} \\
\multicolumn{7}{l}{$^{\rm d}$ Temperature of the Wien seed photon spectrum of \comptt.} \\
\multicolumn{7}{l}{$^{\rm e}$ Powerlaw normalization in units of photons cm$^{-2}$ s$^{-1}$ keV$^{-1}$ at 1 keV.} \\
\multicolumn{7}{l}{$^{\rm f}$ Total photons in the line in units of 10$^{-3}$ cm$^{-2}$ s$^{-1}$.} \\
\label{fit_comptt}
\end{tabular}
\end{center}
\end{table*}

\begin{table*}
\begin{center}
\caption[]{Best-fit parameters of the  \comptb\ model to the analyzed sources. When fitting the source spectra with two \comptb\ components, their electron temperatures $\kte^{\rm tb}$ (thermal plus bulk component) and $\kte^{\rm t}$ (thermal component) are kept equal each other during the fit.
 Emission lines, where observed, are fitted with a simple \gaussian\ model. Parameters within square brackets are kept  frozen in the fit. Parameters within square brackets are kept  frozen in the fit. Photoelectric absorption is computed using the \wabs\ model in \xspec. Errors are computed at 90\% confidence level for a single parameter.}
\begin{tabular}{p{0.13\textwidth}p{0.105\textwidth}p{0.15\textwidth}p{0.15\textwidth}p{0.11\textwidth}p{0.12\textwidth}p{0.12\textwidth}}
\hline
\hline
\noalign{\smallskip}
Parameter          &   Sco~X--1	   &	GX 17+2         &    Cyg~X--2   &   GX~340+0  &  GX 3+1 & GX 354--0 \\
\noalign{\smallskip}
\hline
\noalign{\smallskip}

$N_{\rm H}^{(a)}$ &  [0.15] & 2.10$^{+  0.08}_{-  0.06}$  & 0.26$^{+  0.03}_{-  0.04}$  & 6.35$^{+   0.27}_{-   0.29}$
   &  [1.7] & [3.0] \\
\noalign{\smallskip}
\hline
\noalign{\smallskip}

\multicolumn{7}{c}{\comptb\ (thermal: log(A)=8, $\delta$=0)}    \\
\noalign{\smallskip}
\hline
\noalign{\smallskip}
$\kts^{(b)}$ (keV) & [0.4]  & 0.55$^{+  0.03}_{-  0.03}$   &  0.16$^{+  0.04}_{-  0.04}$ & 0.35$^{+  0.07}_{-  0.07}$ & 0.61$^{+  0.10}_{-  0.14}$  & 1.20$^{+  0.05}_{-  0.05}$ \\
\noalign{\smallskip}

$\kte^{\rm t}$ (keV) &   2.66$^{+  0.06}_{-  0.04}$    &    3.38$^{+  0.07}_{-  0.09}$     &  2.74$^{+  0.05}_{-  0.04}$   &  2.95$^{+  0.07}_{-  0.05}$ &  2.61$^{+  0.11}_{-  0.11}$  & 26$^{+  8}_{- 5}$  \\
\noalign{\smallskip}
$\alpha$ & 0.96$^{+  0.03}_{-  0.04}$  & 0.93$^{+  0.04}_{-  0.04}$  &  0.83$^{+  0.03}_{-  0.02}$    & 0.86$^{+  0.04}_{-  0.03}$  &  1.26$^{+  0.11}_{-  0.09}$   & 1.69$^{+  0.08}_{-  0.08}$  \\
\noalign{\smallskip}
CAF$^{(c)}$ &  2.1      &  2.1      &   3.4   & 2.5   & 1.5  & 1.7 \\

\noalign{\smallskip}
\hline

\noalign{\smallskip}
\multicolumn{7}{c}{\comptb\ (thermal plus bulk)}   \\
\noalign{\smallskip}

\hline
\noalign{\smallskip}

$\kts^{(b)}$ (keV) &  1.21$^{+  0.14}_{-  0.10}$   &  1.45$^{+  0.30}_{-  0.26}$   & 0.96$^{+  0.10}_{-  0.13}$   &  1.38$^{+   0.18}_{-   0.15}$  & --  &  -- \\
\noalign{\smallskip}
log(A)  &  -0.47$^{+  0.52}_{-  0.58}$   & 0.28 ($>$-1.00)  &  0.26  ($>$-0.14) &  -0.44$^{+0.54}_{-   0.47}$ & --  &  --\\
\noalign{\smallskip}
$\alpha$  & 3.63$^{+  0.07}_{-  0.04}$  &   2.37 ($< 2.73$)   &  [2.5]       & 1.76$^{+   0.42}_{-   0.55}$ & --  & -- \\
\noalign{\smallskip}
$\delta$  &  $>$63    & 89 ($>$15)  & 60 ($>$27)    &  52 ($>$23) & --  & -- \\
\noalign{\smallskip}
$\kte^{\rm tb}$  &  [=$\kte^{\rm t}$]   &  [=$\kte^{\rm t}$]    &   [=$\kte^{\rm t}$]     & [=$\kte^{\rm t}$]  & --  & -- \\
\noalign{\smallskip}
\hline
\noalign{\smallskip}
\multicolumn{7}{c}{{\sc gaussian}}   \\

\noalign{\smallskip}
\hline
\noalign{\smallskip}
$E_{\rm l}$ (keV) & -- &    6.69$^{+   0.05}_{-   0.05}$   & 6.77$^{+  0.09}_{-  0.08}$   & 6.73$^{+   0.07}_{-   0.10}$ & [6.4]  &  --\\
\noalign{\smallskip}
$\sigma_{\rm l}$ (keV) &   --    & 0.24$^{+   0.10}_{-   0.09}$  & 0.28$^{+  0.18}_{-  0.13}$  & 0.34$^{+  0.26}_{-  0.12}$  &0.95($>$ 0.51)& -- \\
\noalign{\smallskip}
$I_{\rm l}^{(d)}$ &  --  & 5.5$^{+ 1.3}_{-   1.0}$   &  2.5$^{+  1.0}_{-  0.7}$    & 4.3$^{+2.19}_{-0.90}$  &  11$^{+  4}_{-  5}$  &  --\\
\noalign{\smallskip}

$EW_{\rm l}$ (eV)  &   -- & 47$^{+3}_{-14} $  & 39$^{+18}_{-9}$     &  62$^{+37}_{-10}$   &  286$^{+92}_{-135}$  &  --\\

\noalign{\smallskip}
\noalign{\smallskip}
\hline
\noalign{\smallskip}

L$_{\rm tot}^{(e)} \times 10^{38}$ erg~s$^{-1}$ & 3.65 & 1.53   &   0.96      &  1.42     & 0.19  & 0.06 \\
\noalign{\smallskip}
L$_{\rm comptb,th}^{(e)}$/L$_{\rm tot}$  & 0.79 & 0.86    &   0.74      &   0.90   & --  &  --\\

\noalign{\smallskip}
\hline
${\chi^{2}}$/dof & 214/165 &  158/143 &    189/152     & 182/136   & 48/42  & 155/131  \\

\noalign{\smallskip}
\hline
\noalign{\smallskip}

\multicolumn{7}{l}{$^{a}$ Equivalent hydrogen column in units of 10$^{22}$ cm$^{-2}$.} \\
\multicolumn{7}{l}{$^{b}$ Temperature of the \comptb\ blackbody seed photon spectrum (same as $kT_{\rm bb}$ in Tab. \ref{fit_comptt}).} \\
\multicolumn{7}{l}{$^{c}$ Compton amplification factor, see text for definition.}\\
\multicolumn{7}{l}{$^{d}$ Total photons in the line in units of 10$^{-3}$ \cm2 \s1.} \\
\multicolumn{7}{l}{$^{e}$ Estimated in the energy range 0.1-200\,keV.}\\
\label{fit_comptb}
\end{tabular}
\end{center}
\end{table*}

\section{Application of the model to a sample of LMXBs}
\label{sources}

In order to test our model, we used a data set from a sample of six LMXBs belonging to the Z and atoll classes (Hasinger \& van der Klis 1989): sources of the former class are \mbox{Sco X--1}, \mbox{GX~17$+$2}, \mbox{Cyg~X--2} and \mbox{GX~340$+$0}, while \mbox{GX 354--0} (4U~1728--34) and \mbox{GX 3+1} belong to the latter one. 

The choice of the first four sources was mainly motivated by the fact that they show transient \pl-like hard ($\ga$ 30\,keV) \mbox{X-ray}
emission over their stable continuum (\mbox{Sco X--1}: D'Amico et al. 2001, P06; \mbox{GX~17$+$2}: Di Salvo et al. 2000b, F05, F07; Cyg~X--2: Di Salvo et al. 2002, hereafter DS02; GX~340$+$0: Lavagetto et al. 2004, hereafter L04).
Other sources, such as GX 5-1 (Asai et al. 1994, Paizis et al. 2005), GX 349+2 (Di Salvo et al. 2001) and, most recently, \mbox{GX 13+1} (P06) have shown the same high-energy transient behaviour.
The stable continuum spectra of these source are usually fitted with a two-component model consisting of a
photoelectrically-absorbed \bb\ plus the TC model \comptt\ by T94; when the hard \mbox{X-ray} tail
appears, an additional \pl\  is  included in the model to fit to the data. 
The presence of a gaussian emission line around
6.7\,keV is also observed in these sources.

The persistent continuum spectrum of the bursting atoll \mbox{GX 3+1} is very similar to that of classical
Z sources,  with a Comptonization spectrum characterized by $\kte \sim$ 3\,keV and $\tau \sim$ 10 (Oosterbroek et al.
2001, P06) plus contribution from a $\sim$ 1\,keV \bb-like component. 
Neither spectral transitions nor hard \mbox{X-ray} tails have been observed up to now in this source.

Finally, the bursting atoll source \mbox{GX 354--0} is characterized by the presence of different spectral states that correlate with its position on the hardness-intensity diagram (HID).
Long-term monitoring performed with \integral\ by F06 has shown that in the
lower {\it island state}  of the HID, the source spectrum is quite hard and can be described by a TC spectrum
with $\kte \sim$ 30\,keV and optical depth $\tau \sim$ 1.5  (assuming slab geometry) with no evidence of soft
\bb-like emission. As the source moves across the HID, reaching the {\it banana state}, the TC spectrum gets
softer, with electron temperature dropping to $\sim$ 3\,keV (with $\tau$ correspondingly increasing to $\sim$ 5)
and simultaneous appearance of a soft ($\sim$ 0.6 keV) \bb-like  component which was attributed by F06 to the accretion disk. 

Selection of a sample of sources like these thus allows us to test our model over a wide range of spectral states which are
experienced by NS LMXBs, as it is also  noticed by P06.
Following the  state terminology of P06, we can apply our model to the \emph{intermediate 
state} spectra (\mbox{Sco X--1}, \mbox{GX~17$+$2}, \mbox{Cyg~X--2} and \mbox{GX~340$+$0}) where the combined effect of the thermal and bulk Comptonization is suggested to be the origin of the hard X-ray emission,  to  \emph{very soft} state spectra  (\mbox{GX 3+1}), where thermal Comptonization  by a relatively cold plasma   ($\kte \sim3$ keV) dominates, and  to  \emph{low/hard} state spectra (GX~354$-$0), in which 
the  thermal Comptonization  component characterized by relatively hot electron temperature  ($\kte \sim$ 30\,keV) dominates.

We refer the reader to  P06  for a more complete discussion on 
the interpretation of  the evolution among these spectral states.
It is worth noting that the fit of the fourth spectral state identified by P06 
(\emph{hard/\pl~state}) with \comptb~(similarly to \comptt) will \emph{not} provide
physically meaningful parameters, due to the presence of the non-attenuated \pl~that results
in very high plasma temperatures (hundreds of keV), i.e. out of the regime of validity of the model. 
The spectral shape can be, in fact,  fitted, but the resulting parameters are not physically meaningful (see discussion in Section
4.3).

As a complement to the interpretation by P06, in their analysis of \mbox{GX~17$+$2}, 
F07 proposed an accretion scenario in which the \bb-like 
component in the \mbox{X-ray} spectrum arises from the NS surface, 
while the TC spectrum originates 
in the outer part of the TL region. 
The \pl-like hard \mbox{X-ray} emission is instead  the result of bulk Comptonization of 
the \bb-like (NS) seed photons by the
inward in-falling material in the innermost part of the TL (see Fig.~\ref{geometry}).

In the analysis of the  \emph{intermediate state} sources, we  proceed in the following way: 
first we fit their spectra with a mostly used multi-component  model consisting of a \bb\ plus 
\comptt\ plus a \pl\ plus, when required, \gaussian\ emission lines (see above).
Given that the radial extension of the TL is small compared to its vertical height-scale
(e.g., Titarchuk \& Osherovich 1999, TF08), so that it can be somewhat approximated to a geometrically thick equatorial belt around the NS, we assumed a slab geometry for \comptt\ for all the sources but GX~354--0, whose
spectral state seems to be more related to a quasi-spherical accretion.
Using this multi-component model we do not report the unabsorbed bolometric (0.1-200 keV) source fluxes because the presence of the \pl\ component gives origin to a strong bias at low energies, where in fact the flux diverges, and this effect gets worse as the \pl\ steepness increases. It is worth pointing out this  problem when one  fits  spectra with \pl\ components.

Subsequently, we replace \comptt\ with \comptb, fixing the illuminating factor to $A \gg 1$ 
(i.e. $\log(A)$=8, corresponding to no direct seed photons emission, see Eq. [\ref{flux}]) 
and $\delta$=0 (no bulk contribution); hereafter we call this  \comptb~as \emph{thermal} \comptb.
A  second \comptb\ is used instead of the \bb\ \emph{and} 
\pl~components, letting, this time, $\log(A)$ and
$\delta$ free to vary during the fit (Table~2). The main idea is that the first \comptb\ should just reproduce the TC spectrum 
from the outer TL and presumably most of a Comptonized disk component, while the second one  takes into account the  
TC and BC of the NS surface seed photons, which produce the observed  \pl-like hard \mbox{X-ray} tail.
For the \emph{low/hard state} spectra, GX~354--0, and the \emph{high/soft state} spectra, 
\mbox{GX 3+1}, that  are characterized by pure thermal Comptonization, 
the \comptt\ model was replaced
by a single thermal \comptb~(Table~2) with no need for the second, bulk-related, \comptb~component.
For all of the sources, the optical depth $\tau$ of the TC region is later estimated using the \comptb\ best-fit 
values $\alpha$ and $\kte$ and equations (17), (24) in TL95 and then  compared to the value 
reported from \comptt~(see Table \ref{fit_comptt}).
\newpage
\subsection{Sco~X--1}
\label{scox-1}

The spectrum of \mbox{Sco X--1} was obtained  from observations of the \integral\ satellite (Winkler et al. 2003) 
using data from the Joint European \mbox{X-ray} Monitor (JEM-X, Lund et al. 2003) and the 
high-energy imager 
IBIS/ISGRI (Ubertini et al. 2003, Lebrun et al. 2003).
The ISGRI spectrum was obtained by summation of  all the \integral\ observations of the source 
in the period 2003-2005 (see P06), while for JEM-X, because of  the very high brightness of 
the source (which may give rise to problems in the standard observing mode of the instrument) 
we considered a single pointing ($T_{\rm exp} \sim$ 1800 sec).
We thus caution the reader that the high-energy source spectrum ($\ga$ 20\,keV) shows, in fact, the 
\emph{average} behaviour of the source. 

The inclusion of the JEM-X data at lower energies is  very important in order to get a wide
energy coverage, which is of key importance to constrain the model parameters. Using these two instruments we
can indeed analyze the source spectrum in the energy range 4--150\,keV.
The lack of data below 4\,keV however, does not allow \xspec\ to determine the photoelectric interstellar absorption $\nh$ along the source direction, so we 
fixed its value to 0.19$\times 10^{22}$ cm$^{-2}$,
as found by the Solid-State Spectrometer on-board the {\it Einstein Observatory} (Christian \& Swank 1997).
A simple \comptt\ plus \pl\ model formally would not be  completely acceptable from 
a statistical point of view (\chiq/dof=212/167), even though no systematic deviation are observed in the residuals between the data and the model. Moreover, this result is obtained adding a systematic error of 1\% 
to both ISGRI and JEM-X spectra, which is likely below the current uncertainty on the instrument calibrations. If systematics are increased to 2\%, we obtain \chiq/dof=189/167.
The best-fit parameters of the \comptt\ plus \pl\ model are reported in Table \ref{fit_comptt}.

We subsequently use a two-\comptb\ model: in this case, we deal with two seed photon temperatures,
which can not be determined simultaneously, because the JEM-X low-energy threshold ($\sim$ 3 keV) is
well above the peak of the softer seed photon temperature usually observed in LMXBs ($\sim$ 0.5 keV, see Barret 2001).

We thus fix $\kts$ related to the thermal \comptb\ to 0.4\,keV, allowing instead to vary the seed photon temperature $\kts$ 
suggested to be related to the NS surface.
The results of the fit with a two-\comptb\ model are reported in Table \ref{fit_comptb}: from the best-fit values $\alpha$ and $\kte$ of the thermal \comptb\ component and equations (17) and (24) in TL95, we estimate $\tau$=5.6, which is not far away from that directly find by \comptt\ ($\sim$ 5, see Table \ref{fit_comptt}). 
As it concerns the second \comptb\ component, aimed to describe the hard \mbox{X-ray} tail,  we find that the bulk parameter 
$\delta$  would  be pushed  up to unreasonably high values by \xspec, with no apparent minimum in the \chiq\ space. It is, in fact, just possible to put a lower limit on $\delta$, as reported in Table \ref{fit_comptb}, whereas
good constrains are instead obtained for the seed photon temperature $kT_s$ and the energy index $\alpha$. 
In Figure \ref{efe} we report the deconvolved unabsorbed source spectrum and best-fit two-\comptb~model.
For estimating the source 0.1--200\,keV luminosity we assumed a distance of 2.8\,kpc (Bradshaw et al. 1999).

\subsection{\mbox{GX~17$+$2}}

For GX~17$+$2 we use \sax~(Boella et al. 1997a) observations performed on April 3, 1997, when the source was in the left part of
the horizontal branch (HB, see Fig. 2 in F07) and a hard \mbox{X-ray} tail was present in the source spectrum. 
The data set is the same previously  used  in the analysis by  F05 and F07. 

Following Di Salvo et al. (2000b) and F05, we fit the 0.4--120\,keV source spectrum with a photo-electrically
absorbed \bb\ plus TC (\comptt) component, plus \pl\ to fit the hard \mbox{X-ray} tail, plus a \gaussian\ emission
line around 6.7\,keV. The best-fit parameters of the model are reported in Table \ref{fit_comptt}.
Assuming a distance from the source of 7.5\,kpc (Penninx et al. 1988) and using  the best-fit  \bb\ temperature and flux, we estimate a  \bb\ radius, $\sim$ 5 km. While keeping in mind all the limitations in making such an estimate (i.e. color to effective temperature correction, possible anisotropy in the emission), the order of magnitude of the \bb\ radius  supports the idea that this component very likely originates close to the NS surface. The \pl\ photon index $\Gamma$ is the same already reported in Di Salvo et al. (2000b) and F05.

In the next step,  we have good constrains on the best-fit parameters of the pure thermal  \comptb\ component using the two-\comptb\ model: they are reported in Table \ref{fit_comptb}. Again, from the thermal $\alpha$ and $\kte$-values, we estimate $\tau$=5.2, close to $\tau \sim$ 6 found by \comptt. 
The  situation is  a bit more critical in the thermal plus bulk related \comptb\ component. 
In particular, for the illuminating factor $\log(A)$ and the bulk parameter $\delta$,  the lower limits (at 90\% confidence level) can be only put by \xspec, while, on the contrary, we obtaine only the  upper limit on the $\alpha$-value. 
These statistical uncertainties are mainly related to the small amount of high-energy ($\ga 30$\,keV) points, as it can be seen in 
Figure \ref{efe}, where we report the deconvolved source spectrum with the two-\comptb\ model.

\subsection{Cyg~X--2}

For Cyg~X--2
we use a \sax~data set   which has been already analyzed  by DS02. These data are related to a source observation performed on 1996 July 23. During this observation, the source traced the full HB across its colour-intensity diagram. The HB was divided into two parts (upper and lower HB, respectively, see Fig. 2 of DS02) for which separated 
spectral analysis was carried out; in both cases a hard \mbox{X-ray} tail was detected. 

In this Paper we report 
results from the upper HB.
DS02 fitted the 0.1-100\,keV continuum with a multi-colour disk blackbody (\diskbb\ in \xspec, Mitsuda 
et al. 1984) plus \comptt\ plus \pl\ plus two \gaussian\ emission lines at $\sim$ 1\,keV and $\sim$ 6.7\,keV, 
respectively.
We note however that DS02 used for the Low-Energy Concentrator Spectrometer (LECS, Parmar et al. 1997) the standard on-line available response matrix. 
The latter may give rise to spurious features when the instrumental count rate exceeds 
$\sim$ 50\,counts~s$^{-1}$. For this \mbox{Cyg~X--2} observation, the LECS count rate is 
$\sim$ 60\,counts~s$^{-1}$, thus requiring the production of an observation-related response function,
which was generated using the {\it LEMAT} package (v5.0.1).

With the newly produced matrix we establish, in fact,  that there is no more strong evidence
of the ~1\,keV emission line. On the other hand, an excess around 2.6\,keV is  still found in the residuals
of both the LECS and the Medium-Energy Concentrator Spectrometer (MECS, Boella et al. 1997b).
We fit this excess with a \gaussian\ emission line, even though it is not clear whether its origin
may be attributed to the source or arise from some instrumental effect. 
The results of the fit with a \bb\ plus \comptt\ plus \pl\ model are reported in Table \ref{fit_comptt}.
The \pl\ index $\Gamma$ results steeper than that reported in DS02. This is mainly due to the different
response matrix used at low energies ($<$ 4\,keV), namely  in the region where the interstellar absorption $\nh$ and
\pl\ normalization must be simultaneously determined by \xspec. We also estimate the \bb\ radius, finding
$\sim$ 8 km, for a source distance of 8\,kpc.

Replacing the above model with the two-\comptb\ one, similarly to \mbox{Sco X--1} and \mbox{GX 17+2}, 
we find very good constraints on the best-fit parameters  related to the thermal \comptb~(see Table \ref{fit_comptb}). 
As in the previous sources, the optical depths inferred from best-fit $\alpha$ and $\kte$-values, and  estimated by \comptt\ are very close each other,
($\tau \sim 6.3$ and $\tau \sim 6.9$, respectively).

As it concerns the thermal plus bulk parameters, unlike \mbox{Sco X--1} and \mbox{GX 17+2}, where, even though poorly constrained, they can be left free in the fit, there is no way to perform serious analysis leaving all the parameters free.
For Cyg X--2 we  thus performed spectral fitting in several steps with $\alpha$ fixed at each step in the range  from 0.5 to 3.5, finding that the minimum in the \chiq\ parameter space is obtained for $\alpha$=2.5.  Fixing $\alpha$, just allows us to put lower limit (90\% confidence level) for $\delta$ and $\log(A)$. 
In Figure~\ref{efe} we report the deconvolved source spectrum with the two-\comptb\ models.

\subsection{GX~340$+$0}

The spectrum of GX~340$+$0 refers to a pointed \sax\ observation of the source performed between 2001 August 9 and 10. A more detailed description of the data reduction can be found in L04.
The results with the \bb\ plus \comptt\ plus \pl\ model, with an additional \gaussian\ emission line, are reported in Table \ref{fit_comptt}.  
It is worth pointing out that the \comptt\ model  with slab geometry gives a slightly worse result with respect to that with spherical geometry (for which we obtain \chiq/dof=160/137).
We also want to emphasize the different values of the  direct \bb\ and \comptt\ seed photons temperature 
($\ktbb$ and  $\ktw$, respectively) reported in Table \ref{fit_comptt} in comparison with the results of L04.
As discussed in F05, the usually adopted model \bb\ plus \comptt\ for description of the persistent continuum 
spectra of LMXBs, may give rise to a dichotomy in the solutions, given that two solutions (statistically 
indistinguishable) are possible, one with $\ktbb > \ktw$ and the other with $\ktbb < \ktw$. In the 
former case, the \bb-like emission is suggested to mainly come from the NS surface, while in the 
latter one it is presumably related to the accretion disk. These different interpretations are 
also suggested by the inferred \bb-radii which are of order of the NS radius in the former case 
or significantly greater ($\sim$ 78 km in L04) in the latter one.
On the basis of the accretion scenario proposed by F07, we prefer the first solution 
(see Table  \ref{fit_comptt}). 

The results with the two-\comptb\ model are instead reported in Table \ref{fit_comptb}.
In this case, the general considerations on the other three sources can be extended to \mbox{GX 340+0}: the constraints
on the best-fit parameters of the thermal \comptb\ component are very good, while they are worse for the 
thermal plus bulk related one, except for the seed photon temperature.
Noticeably, the  inferred $\tau$ from the best-fit (thermal) $\alpha$ and $\kte$ is lower ($\tau\sim6$) than the 
one directly estimated with \comptt\ ($\tau\sim9$, see Discussion).
For the estimation of the bolometric (0.1--200\,keV) source luminosity we assumed a distance of 10\,kpc (L04). 

\subsection{\mbox{GX 3+1}}

We use the data  from the \xte\ (Bradt et al. 1993) public archive related to a source observation performed on August 21 2004.
A joint spectrum from the second units of the Proportional Counter Array (PCA, Jahoda et al. 2006) and
High-Energy \mbox{X-ray} Timing Experiment (HEXTE, Rothschild et al. 1998) on-board the spacecraft was obtained, following the
standard procedure for data reduction. At high energies the source was detected by HEXTE only up to 30\,keV.
Similarly to \mbox{Sco X--1} \integral\ spectra, the low-energy threshold of 3\,keV does not allow us to constrain the photoelectric interstellar absorption $\nh$ in the source direction, 
so we fixed its value to  1.7$\times 10^{22}$  cm$^{-2}$ (Christian \& Swank 1997). 

We find that a simple photoelectrically-absorbed \comptt\ model is good enough to describe the source continuum
spectrum. However, a clear excess in the region 6-7\,keV reveals the presence of a iron emission line which is fitted
with a simple \gaussian. The poor instrumental resolution in this energy region, and the uncertainty calibration
of the instrument because of the diffuse Galactic ridge emission, do not allow  us to simultaneously  constrain
all the line parameters. We thus fix the line energy centroid at 6.4\,keV during the fit. Because of these
uncertainties, both the line broadening $\sigma_l$ and equivalent width values, must be considered with some caution.
The best-fit parameters of the \comptt\ plus \gaussian\ model are reported in Table \ref{fit_comptt}. 

As in the
case of \mbox{Sco X--1} (see \ref{scox-1}), the absence of a direct \bb-like component in the spectrum is very
likely due to a combination of the low-energy PCA threshold (3\,keV) and data quality.
Subsequently, replacing \comptt\ with thermal \comptb, we find the same statistical result, with best-fit 
values reported in Table~\ref{fit_comptb}. 
The inferred optical depth of the TC region, obtained from best-fit $\kte$ and $\alpha$ values, is  $\tau \sim$ 4.9, very close to that found by \comptt\ ($\tau \sim$ 5.3, see Table  \ref{fit_comptt}).
The bolometric 0.1--200\,keV source luminosity is estimated assuming a distance of 5\,kpc. We note that this value is 
certainly underestimated, as it includes the region below 3\,keV, not covered by PCA, where a \bb-like component has been already observed from the source  (Oosterbroek et al. 2001).

\subsection{GX~354--0}
\label{gx354}
The spectral analysis on \mbox{GX 354--0} is carried-out using data obtained by F06 with JEM-X  and ISGRI  onboard the \integral\ satellite. 
Among the nine spectra extracted by F06 as a function of the source position in the HID,
we consider the hardest one (see Fig. 3 and Table  2 in F06). Unlike F06 however, we assume a spherical geometry for the plasma in \comptt, given that it seems to be more representative for the  hard state accretion scenario.  
A simple photoelectrically-absorbed \comptt\  model is  good enough to fit the source spectrum and no soft \mbox{X-ray} emission is seen in the spectrum.
 
Replacing \comptt\ with a thermal \comptb, we obtain the same results of \comptt, both in terms of statistics and as it concerns the agreement between the seed photon and electron temperatures (see Tabs. \ref{fit_comptt} and \ref{fit_comptb}).
Using equations (17) and (24) of TL95 for the spherical case, from  the best-fit $\kte$ and $\alpha$-values we derive
$\tau \sim$ 1.5, which perfectly matches the best-fit value directly obtained from  \comptt.
The source bolometric 0.1--200\,keV luminosity was  computed assuming a source distance of 5\,kpc (Di Salvo et al. 2000a; Galloway et al. 2003). Figure \ref{efe} shows the deconvolved source spectrum and superposed best-fit model.

\section{Discussion}
\label{discussion}

We present a new model for the \mbox{X-ray} spectral fitting package 
\xspec, named \comptb. The principal aim of this model is to provide a more
physical description of the transient \pl-like hard \mbox{X-ray}  ($\ga30$\,keV) tails observed 
in LMXBs, in the framework of the bulk motion Comptonization theory.
We recall however  that \comptb\ is actually not just specific to bulk motion. Indeed, setting the bulk parameter
$\delta$ to zero, it reduces to a \emph{generalized} thermal Comptonization model, in the same fashion as the widely
used models \compst\ or \comptt.

\subsection{Thermal Comptonization component}

In the six sources that we analyzed, the  spectra are dominated by a TC component which carries-out
most of the total energetic budget (see Table \ref{fit_comptb} and Fig. \ref{efe}).
In the case of \mbox{GX 3+1} and \mbox{GX 354--0}  this is in fact  the only component observed in the \mbox{X-ray} spectrum.

In the  GX 3+1 case, this is a result of a low-energy ($\sim$ 3 keV) threshold effect of PCA, given
that a \bb-like emission has been  previously observed with \sax\ (Oosterbroek et al. 2001).
On the other hand, in the case of GX 354--0,  it is  possible that this soft emission was \emph{intrinsically} absent or very weak
in the analyzed  spectrum of the source given that it was  detected  when  the source  was in the intermediate   state
(see  F06  and also Di Salvo et al. 2000a).
 F06 attributed the appearance  of the soft \bb-like emission, in the soft state,  to the accretion disk.
Instead, in our accretion scenario (as already reported in F05), 
the \bb-like emission is related  to the emission from the  NS star surface.
Independently of the emission  region from which it comes, the fact that this component is not
observed in the \emph{low/hard}  state spectrum of GX~354--0 is consistent with an accretion geometry where the TC corona is significantly extended and  being  almost spherical.
This spherical  configuration should completely cover the central object, intercepting all the seed photons coming
from the NS surface.

For all of the studied sources, we compute the \emph {Compton amplification factor} (CAF)  defined as the ratio of the Comptonized to seed photon energy fluxes. 
Noticeably, the  \emph {Compton amplification factor} (CAF)  gets  the highest value in  \mbox{Cyg X--2} (3.5), despite the fact that its TC component provides the lowest fractional
contribution to the total spectrum among the six studied sources (see Table \ref{fit_comptb}).
This is ultimately related to the fact that the seed photon energy is softer ($\kts \sim$ 0.2) in \mbox{Cyg X--2} than in the remaining five sources (see Chakrabarti \& Titarchuk 1995, for the CAF dependence  on the seed photon energy). 

It is important to emphasize that  the energy index $\alpha$  directly provides the efficiency of the Comptonization process (see e.g. Bradshaw et al. 2007, hereafter BTK07), no matter which is the plasma geometry (i.e. slab or  sphere). For a given geometry  (taking into account  hydrodynamical considerations), one can infer  Thomson optical depth $\tau$ using  the
best-fit $\alpha$ and $\kte$ values provided by \comptb\ and using equations (17) and (24) in TL95.

We find that the derived  values are in very good agreement with those obtained from \comptt, except for
 \mbox{GX 340+0}, where $\tau$ provided from \comptt\ ($\sim$ 9.2) is higher than that estimated
using \comptb~($\sim$ 5.8).  This difference likely comes  from the slightly different  description of the Comptonization by these two models.

\subsection{Combined thermal-bulk Comptonization component}

The bulk motion Comptonization process as a possible origin of hard \mbox{X-ray} tails
in LMXB systems hosting either a NS or a BH has been suggested from both observational (Shrader \& Titarchuk 1998;
Borozdin et al. 1999; P06, F07) and theoretical
(TMK96, TMK97, Laurent  \& Titarchuk  1999, 2001; TF08) works.
A model which can be used to describe these emerging spectra is already present in the \xspec\ package
(\bmc, TMK97), but as previously discussed in Section \ref{introduction}, it presents some limitations, which  are the lack of the electron recoil term (cutoff) in the Green's  function and the parameterization of the high-energy Comptonized component in terms 
of just the spectral index $\alpha$ \textbf(no $\delta$ bulk parameter). 
In fact, index $\alpha$  has  only information about the efficiency of Comptonization, no matter whether  it is purely thermal or due to combined effects of thermal and bulk Comptonization (see TMK97 and BTK07).

We try to overcome these limitations,  including the full  Green's  function expression 
(Eq. \ref{green})  in our model  and using numerical convolution of the Green's
function with the seed photon spectrum (Eq. \ref{convolution}).

 One of the advantages of this numerical approach is that the condition $\kts \ll E_{pl}$ (seed photons energy much 
less than plasma energy  and Green's function approximated by a broken \pl, see ST80, TMK97) is not required. 
This model modification is, in fact, important in the spectra of high-luminosity NS LMXBs, where the seed photon and plasma energy (electron temperature) differ just for a factor of a few (see Tables \ref{fit_comptt} and \ref{fit_comptb}).

We apply our combined thermal plus bulk  Comptonization model ($\delta > 0$) to the four LMXBs of
our sample  which shows a \pl-like hard \mbox{X-ray} tail, namely in \mbox{Sco X--1}, \mbox{GX 17+2}, \mbox{Cyg X--2} and
\mbox{GX 340+0} spectra.  In particular, we use   a two-\comptb\ model for these sources  in which  the first \comptb~
takes into account the pure TC part of the spectrum (see previous subsection) while the second \comptb~describes 
both the \bb-like emission observed at low energies and the \pl-like emission above 30\,keV. 
We  assumed the same  temperature $\kte$ for the thermal and thermal plus bulk Comptonization region. The correctness of this assumption depends on how the two regions are physically and geometrically coupled and whether an actual temperature gradiend does exist
on the region where Comptonization occurs.

In the accretion geometry proposed by F07, bulk Comptonization occurs in the innermost part of the transition
layer region, while TC is  dominant in the outer TL and presumably in some extended region located above  the accretion disk.
The latter statement is suggested by the lack of a direct soft ($\sim$ 0.5\,keV) \bb-like \mbox{X-ray} emission in the spectra, usually attributed to the accretion disk. 
It is  likely that most of the disk emission is up-scattered in the TL or somewhat embedded in the total
spectrum, without possibility to split it as a single spectral component.
We find some indication of contribution of disk soft seed photons to the TC component using  the best-fit values of the BB temperature of
these photons. Note  this temperature is significantly lower than that of the \bb-like seed photon component presumably   coming  from   NS surface
(see   Tabs. \ref{fit_comptt} and  \ref{fit_comptb} for \comptt\ and \comptb, respectively).
Particularly for Cyg X--2 we obtain a value $\la$ 0.2 keV, in  contrast  to higher values of that found  by DS02 or by Done \& Gierlinski (2003).
Higher disk temperatures are also found by Gilfanov et al. (2003) and Revnivtsev \& Gilfanov (2006). 

However we should point out  that  all these authors  (but DS02) used PCA/RXTE data for which the low-energy band is higher than 3 keV,  whereas our best-fit  seed photon temperatures are based on the \sax\  spectra in the broad energy band from 0.4 to 120 keV. 
The PCA/RXTE with its low-energy threshold above 3 keV cannot in principle  determine any trace  of the soft component whose temperature is lower than 1 keV given that  its maximum emission ($\la$ 3 keV) would still fall outside the PCA energy band.
On the other hand the disk photon temperatures of 0.8-1.4 keV found by DS02   is inferred  using a response matrix for the  low-energy instrument (LECS) which is not correct for such bright sources and thus (as also testified by strong decrease of the 1 keV emission line in our data) must be considered very cautiously.

 The uncertainties of the bulk-related parameters of \comptb\ (see Table  \ref{fit_comptb})
do not allow us to check  the actual differences among them as it concerns their innermost bulk-dominated region given  the presence of  a few points only  at high-energies ($\ga$ 30 keV) in the spectra of the four \emph{intermediate state} studied sources
(see four upper panels in Fig. \ref{efe}).

However we should note  that in all cases $\delta$, through which the relative importance of BC to TC is parametrized, is of order of tens, which values are  expected in the case of   efficient BC,
when the  hard \pl-like emission should be  seen.
The fact that we can  not constrain  the upper value (at 90\% confidence level) of $\delta$ is not surprising because of 
 the measurement of the high-energy cutoff (not observed in the analyzed spectra) is required. 

Indeed, the most important effect of the bulk term in the Green's function is  to push forward the rollover energy, with respect to the pure TC case: higher  $\delta$-values lead to higher  energies at which the spectrum exponentially falls down. This can be clearly seen in Figure \ref{green_plot}. On the other hand, it is possible to provide only a lower limit of $\delta$ if  the high-energy behaviour of the available data
goes as an unbroken \pl. This is exactly what happens for the \integral/IBIS spectrum of Sco X-1 (see Fig. \ref{efe}).

In the case of \mbox{GX 17+2}, \mbox{Cyg X--2} and \mbox{GX 340+0}, the thermal plus bulk \comptb\ component predicts a rollover of the spectrum in the energy range 100-200 keV (see Fig. \ref{efe}). However, it is not possible with the present data to really confirm this prediction, even  the fact that \xspec\ finds a minimum of $\delta$ in the \chiq\ parameter space could not be indicative of that.

In the thermal plus bulk \comptb~component, the Green's function energy index $\alpha$ is, for all sources,
higher than that obtained for pure TC, with the highest difference observed in \mbox{Sco X--1}. 
In Figure \ref{green_plot}, we plot the Green's function for monochromatic input line at energy $x_0$=0.3 (where
$x_0 \equiv E_0/\kte$), for two different values of $\alpha$ (1 and 2.5, respectively) considering pure thermal
($\delta$=0) and bulk-dominated ($\delta$=30) cases.  The chosen value of $x_0$ is, in fact, representative of the
NS local environment condition under this  study, as $\kts \sim$ 1\,keV and $\kte \sim$ 3\,keV (see Tables \ref{fit_comptt} and \ref{fit_comptb}). 
Looking at Figure \ref{green_plot}, one can note that high $\delta$-values push forwards the high-energy  cutoff, as already mentioned above.
It is  evident that the higher $\alpha$-values related to the BC region with respect to  those related to TC region (see Table  \ref{fit_comptt}),  are indicative of the fact that in the  innermost part,  for  all four sources, the TC  efficiency  is reduced but the  BC effect  dominates over the TC one given that the power-law hard tail is detected.

\subsection{Limitations and applicability of the model}

When one uses \comptb\ to fit the \mbox{X-ray} spectra of NS LMXBs, there are some issues which must be kept in mind.

Let us first consider the case of pure TC spectra ($\delta$=0 in the Green's function, see TMK97 and  Eq. [\ref{green}] here).
The Green's function of the model is derived as a solution of the diffusion Comptonization equation, which is in turn obtained
as an approximation of the  Boltzmann kinetic equation in the diffusion regime.
This means that the \emph{average} photon energy exchange per scattering is small ($\Delta E/E \ll 1$) and photons suffer many scatterings when traveling across the plasma before escaping.

This condition  holds when the plasma temperature is about 20--30\,keV and less but its optical depth is
high, such as observed in the persistent spectra of Z sources and high-luminosities atoll sources (see Barret
2001 for a review). 
In general  most of the NS LMXBs obey this condition, as the strong radiation flux 
coming from the central object acts as a thermostat in controlling, through Compton cooling, 
the plasma temperature (Barret et al. 2000). 

In the case of BHCs, the situation is quite 
different and typical hard states for this class of sources are
characterized by plasma temperatures of  60 keV  (see e.g. Remillard \& McClintock 2006) and optical depth a few.
In this case, the diffusion approximation should be used with some care (the photon energy exchange for scattering is not so small but   the number of scatterings before photons escape is relatively small), and a different treatment of the Compton 
scattering should  be addressed: models such as \compps\ (Poutanen \& Svensson 1996), which use iterative
scattering numerical procedures are thus most suitable to be used for fitting BHCs hard state spectra.
This is one of the reasons why \comptb\ (but also \compst\ or \comptt) should be applied preferably 
to NS LMXBs spectra with plasma temperatures below 20--30\,keV. 
In  other cases, BHs or NS in their \emph{hard/\pl~state} (see Section 3), 
the model can just \emph{fit the shape} of the spectrum: while this may result 
satisfactory in terms of $\chi^2$ value, the output parameters should be taken cautiously.

Let us now concentrate on the case where BC dominates over TC ($\delta \gg$1).
The analytical  solution of the Comptonization equation (\ref{kompaneets}), for the Green's function,  does exist only neglecting the term, $({v_{\rm b}}/c)^2/(3\Theta)$ in equation (\ref{kompaneets}), where $\Theta \equiv kT_{\rm e}/m_{\rm e}c^2$. 
Whether this assumption is correct enough, it  depends on the local environment conditions close to the central object.
In the case of systems hosting a NS, the presence of a firm surface plays a key role in determining the global
inward motion of the accreting material. At high accretion rates, the NS surface on  one side is the source of
a strong radiation pressure due to kinetic energy release of the matter, but on  the other hand  its mirror-like
behaviour (reflecting inner boundary condition) is crucial in determining, at any radius, the pressure gradient due
to local gravitational  energy release of the in-falling material. 

In many cases, these concurring effects may efficiently stop (or strongly decrease) the in-falling matter velocity $v_{\rm b}$,
thus suppressing the bulk Comptonization effect (and associated hard \mbox{X-ray} tail). 
Numerical simulations by TF08 show that under this conditions $v_{\rm b}$ is $\la$ 0.2~c, and typical
in-falling velocities close to the NS surface are $\sim$ 0.1~c.
In this case, for an electron temperature $\kte$=3 keV, it results
$({v_{\rm b}}/c)^2/(3\Theta)$=0.6.

In the case of BHCs, the different inner boundary condition at the event horizon (e.g, Titarchuk \& Fiorito 2004)
strongly modifies the radiation pressure behaviour with respect to the NS case, and may lead the inward bulk velocity $v_{\rm b}$ to get values
of $\sim$ 0.3-0.4 $c$ in proximity of the last stable marginal orbit, even at accretion rates higher than
Eddington limit. Given that the typical temperatures for these states are similar to those of the NS \emph{ intermediate} and 
\emph{very soft} states (e.g., Titarchuk \& Fiorito 2004), namely $\sim$ 3-5 keV, we derive that  $({v_{\rm b}}/c)^2/(3\Theta)$ may 
reach a few.  The full kinetic treatment  should  be used to determine the Green's function for this case (see  Titarchuk \& Zannias 1998, Laurent \& Titarchuk 1999) because  the value of $v_{\rm b}\gax 0.3c$ and the photon energy change is not small in the bulk flow into BH.

The numerical solution of the Comptonization  equation including  the second order bulk term shows that 
it produces two net effects on the the emerging spectrum: it pushes  forward the 
high-energy cutoff and makes higher the Comptonized spectrum normalization (see Fig. 1 in TMK97). 
In terms of our model parameterization, given that the high-energy cutoff is not observed in 
the data, we would expect that this higher Compton normalization, actually due to $(v_{\rm b}/c)^2$, 
would be somewhat compensated by a higher illuminating factor $\log(A)$ value.
The model can thus always \emph{fit the shape} of the spectrum, whereas its best-fit parameters should be taken with some precaution 
as in the case of pure thermal, high-$\kte$, TC spectra (see above).

A second issue to be pointed out is how one should define $\delta$. The analytic form reported in equation (\ref{delta}) is derived for a plasma subjected to a pure free-fall radiation-corrected velocity profile, where only gravity and radiation pressure from the central object are the concurring processes.  In particular, if the compact object is a BH, only gravity is present.

The condition where only gravity and radiation pressure are at work holds, in fact,  in the case of a \emph{optically thin} medium, where the radiation flux (from the NS) follows the  $R^{-2}$ law.
However, it is known that high-luminosity LMXBs systems which show hard \mbox{X-ray} tails (the six known Z sources and, recently, also \mbox{GX 13+1}, see P06) are characterized by high $\dot{M}$ values (mainly inferred from their luminosity), and consequently optically thick environments.
In this case, additional (and even dominant)  effects due to radiation pressure from local gravitational and kinetic energy releases, viscous energy transportation, gas and  magnetic pressure, cannot be neglected. The numerical simulations performed by TF08 show  that under these conditions the radial velocity behaviour of the in-falling material deviates from a pure free-fall law across the {\it whole} TL. Nevertheless, in the innermost part of the TL itself, gravity and radiation pressure actually dominate and the velocity profile is almost  free-fall radiation-corrected like and $\delta$ definition as defined in
equation  (\ref{delta}) can be almost safely used.

To strengthen this statement, we tried to make a independent $\delta$-estimation as it would be derived in the optically thin case (see Appendix).
It is worth noting that in equation (\ref{delta_estimation}), the only parameter which cannot be directly obtained  from the
observations is the dimensionless accretion rate $\dot{m}$. We find the $\delta$-values 155/$\dot{m}$, 145/$\dot{m}$, 178/$\dot{m}$ and 169/$\dot{m}$ for \mbox{Sco X--1}, \mbox{GX 17+2}, \mbox{Cyg X--2} and \mbox{GX 340+0}, respectively.
While keeping in mind the issues discussed above,   the similarity of the $\delta$-values, as order of magnitude, between  these values and those reported in Table \ref{fit_comptb} is interesting. Higher values of  $\dot{m} > 1$ would even reduce this difference.

\subsection{Similarities and differences between \comptb\ and BW07 model}

As already pointed out in the Introduction, BW07 developed a TC and BC model for accreting X-ray pulsars.
The radiative transfer form reported in equation (34) of BW07 is  sligthly different from that reported in equation (14) of TMK97, which we adopted in our Paper (see Eq. [\ref{kompaneets}]). 
The difference mainly arises because of the different assumptions on the velocity profiles in TMK97 and BW07.

 A free-fall velocity profile $v(r) \propto r^{-1/2}$ is used by TMK97 which leads  to  $\tau(r) \propto r^{-1/2}$ and  $v(\tau) \propto \tau$ using the continuity equation 
 (see TMK97 and also TMK96). 
 This result is obtained keeping in mind that the surface through which matter flows scales as $r^2$.

Instead,  in the model of BW07, plasma motion channeled by the strong magnetic field is modelized as a flow passing  through a cylindric column whose area does not change with the distance $r$ from the NS polar cap (see their Eq. [19]).
Using this particular form of the continuity equation BW07 show that  $\tau(r) \propto r^{1/2}$.
 In order to analytically treat (by mean of variable separation) the problem, BW07 assume that $v(\tau) \propto \tau$ and thus $v(\tau) \propto r^{1/2}$.
This direct, instead of inverse, proportionality between $v(r)$, $\tau(r)$ and $r$ would obviously modify the form of the spatial term in the Comptonization equation when writing it as a function of the optical depth $\tau$ instead of the distance $r$.

Moreover we note that in BW07 optical depth $\tau$ is  defined as such that $d\tau$=$N_{\rm e} \sigma_{||} dr$ (where $\sigma_{\rm ||}$ is the Thomson cross-section for photons propagating in parallel direction to the magnetic field lines and $r$ is a radial length coordinate), while TMK97 use an \emph{effective} optical depth, defined such that $\tau_{\rm eff} (r_{\rm trap})$=1, where $r_{\rm trap}$ is the photon \emph{trapping radius}. 
Looking at equations (1) and (5) in TMK97 is evident that with such a parametrization  $\tau_{\rm eff}(r) \propto r^{-1/2}$ but $v(\tau_{\rm eff}) \propto \tau_{\rm eff}^{1/2}$.

It is worth noting that also BW07 neglect the second order velocity $(v_{\rm b}/c)^2$ bulk term, which is essential in order to separate Comptonization equation in space and energy.
Another important difference between the two models is that  BW07  study the BC effect in  the presence of a strong magnetic field ($B \sim 10^{12}$ G) which modifies the electron cross section in the photon direction parallel and perpendicular to the magnetic field lines.
However it  is not our case, as the class of sources we consider are characterized by  B-fields of order B$\la 10^8$ G and less (see Titarchuk et al. 2001, for details of B-field determination in LMXBs).  
For such low magnetic fields the B-field modification of the electron cross-section can be neglected.

However it is remarkable that for both models the Green's function  is a broken \pl\ in the  energy band where the recoil effect can be neglected (compare Fig. \ref{green} in our Paper and Fig. 5 in BW07). Also it is common for these two models that the high-energy \pl\ is followed by exponential turnover when the recoil effect is taken into account. It implies that the shape of the emergent spectrum  is similar  and it can be also fitted  by  the \xspec\ models \comptt\ and \bmc. The main difference is how the \pl\ index of the Green's function and high-energy cutoff are related to the physical parameters of the particular model. 
We can consider our models to be somewhat complementary as, starting from the same physical process but in very different enviroments, they are supposed  to describe emergent spectra for two classes of NS binary systems.

\section{Conclusions}
\label{conclusion}

We developed a new model for the \mbox{X-ray} spectral fitting package \xspec\ (\comptb) which  can 
be considered a general Comptonization model and more specifically the extension of the 
previously developed Bulk Comptonization Model (\bmc), already  present in \xspec. 
Similarly to \bmc, our model consists of two components: one represents the \bb-like 
seed photons which escape the plasma cloud without appreciable energy exchange, the other one gives the 
\emph{effectively} Comptonized spectrum, obtained as the
convolution of the system Green's function with the seed input photon spectrum.

Using spectra of the  \sax, \integral\ and \xte\  satellites, we tested the model on a sample 
of six NS LMXBs which experience different spectral states: \emph{intermediate state} 
(\mbox{Sco~X--1}, \mbox{GX~17$+$2}, \mbox{Cyg~X--2} and \mbox{GX~340$+$0}), 
where a \pl-like hard \mbox{X-ray} component appears over the 
stable continuum dominated by  thermal Comptonization, low-$\kte$, high-$\tau$ spectrum, a \emph{very soft state} 
(\mbox{GX~3$+$1}), similar to the previous one but without the \pl-like component, and a \emph{low/hard state} 
(GX~354--0) where only thermal Comptonization from a hot plasma ($\kte \sim$ 30 keV, $\tau \sim$ 1.5)  is observed in the spectrum.

We find an excellent agreement, as it concerns fitting the TC components,  between the best-fit parameters 
of \comptb\ and
those of \comptt\ (the most widely used model to this scope), which confirms the goodness of the code.

On the other hand, one of the most important goals of \comptb\ is to try a more physical approach,  in the 
framework of the bulk motion Comptonization process, in explaining  the transient \pl-like 
hard \mbox{X-ray}  ($\ga30$\,keV)  emission observed
in NS LMXBs. This is what we have done in analyzing the spectra of the four intermediate state sources.
It is worth noting that bulk motion Comptonization is supposed to be at work, with even 
higher efficiency, also in BH systems in their soft states,
where a unbroken \pl\ extended up to MeV energy is usually observed.

We show that the values of the  $\delta$-parameter, which represents the importance of bulk with respect to thermal  Comptonization, obtained from the fit, can be physically meaningful and can quantitatively describe the physical conditions of the environment in the innermost  part of the NS systems. 
For three of the \emph{intermediate state} sources (\mbox{GX 17+2}, \mbox{Cyg X--2} and \mbox{GX 340+0}) the best-fit model 
 predicts a high-energy cutoff in the spectrum in the energy range 100-200 keV, which is however not possible to confirm on the basis of the present available data.

The next generation of high-energy missions, with their improved sensitivity and extended energy coverage, will definitively address  this issue. On the other hand, the theoretical results coming from a full magneto-hydrodynamical treatment of the TL region in NS and BH systems  are very promising. 
 We are planning to perform Monte Carlo simulations where
 the derived radial velocity profiles for NS systems  will be of key importance in order to predict the position of the high-energy cutoff due to BC effect, to be then compared with future observations.

\begin{acknowledgements}
The authors thank G. Lavagetto and M. Falanga for providing their data on the spectra
of GX~340$+$0 and GX~3$+$1. 
AP acknowledges the Italian Space Agency financial and 
programmatic support via contract I/008/07/0.

\end{acknowledgements}

\appendix
\section{Estimation of the $\delta$-parameter from observable quantities}

\noindent
The $\delta$-parameter for an optically thin accreting plasma is given by
\begin{equation}
\delta=\displaystyle{\sqrt{1-\ell} \over {\dot{m}\Theta}},
\label{delta_a}
\end{equation}
where $\ell \equiv F(R)/F_{Edd}(R)$ is the ratio of the actual to Eddington fluxes at any distance R,
$\dot{m}$ is the accretion rate in Eddington units and $\Theta$ is the dimensionless electron temperature
$\Theta \equiv \kte/m_{\rm e} c^2$.
In the case of isothermal plasma, given that  both F(R) and $F_{\rm Edd}$(R) go as $R^{-2}$, the parameter  $\delta$ is  independent of R.
Let us define now $F_{\rm obs}$ as the flux measured by the observer at distance D. This is related to the flux at a given
distance R by 
\begin{equation}
F(R)=\displaystyle{D^2 \over R^2 } F_{\rm obs}.
\label{flux_r}
\end{equation}
\noindent
Keeping in mind the definition of the  Eddington flux $F_{\rm Edd}$=$GMc\mu_{\rm e}/kR^2$  (where $k$=0.4 cm$^{2}$ g$^{-1}$ and $\mu_{\rm e}$
is the average electron molecular weight, which we assume equal to 1.14 as for solar abundances) and using equation (\ref{flux_r})
 we may rewrite equation (\ref{delta_a}) as 

\begin{equation}
\delta=\left(1-  \frac{F_{\rm obs}D^2 k}{GMc\mu_{\rm e}} \right)^{1/2}\displaystyle{ 1 \over {\dot{m}\Theta}}.
\label{delta_estimation}
\end{equation}
\noindent
If we relate $F_{\rm obs}$ to the \bb-like flux coming from the NS, then we  can estimate $\delta$ using the best-fit \comptb\ parameters, by setting to zero the illuminating factor A, or correspondingly, log(A)$\ll$1. This would indeed provide just the flux of the direct \bb-like component of the spectrum 
(see Eq. \ref{flux}).
On the other hand the dimensionless  parameter $\Theta$ can  be estimated using the best-fit temperature $\kte$.  
Thus, assuming a NS mass $M$=1.4$~M_{\odot}$ and if the distance $D$ from the source is known, it is possible to estimate
$\delta$ as a function of the  mass accretion rate $\dot{m}$, which is the only quantity not  directly measurable from the observations.

\clearpage

\end{document}